\documentclass[fleqn,usenatbib]{mnras}

\usepackage{lipsum}
\usepackage{mwe}

\usepackage{lipsum,graphicx}
\usepackage{newtxtext,newtxmath}
\usepackage[T1]{fontenc}
\usepackage{ae,aecompl}
\usepackage{adjustbox}
\usepackage{amsfonts}
\usepackage{amsmath}
\usepackage{graphicx}	
\usepackage{tabularx}
\usepackage{verbatim}
\usepackage{xspace}
\usepackage{adjustbox}
\usepackage{rotating, lipsum, babel}

\usepackage{subfig}

\newcommand{\pc}{\,\mathrm{pc}}
\newcommand{\Msun}{\,\mathrm{M}_{\odot}}
\newcommand{\kpc}{\,\mathrm{kpc}}
\newcommand{\Gyr}{\,\mathrm{Gyr}}
\newcommand{\Gyrs}{\,\mathrm{Gyrs}}
\newcommand{\kms}{\,\mathrm{km\,s}^{-1}}

\newcommand{\RG}{R_\mathrm{G}}

\newcommand{\Mb}{M_\mathrm{b}}
\newcommand{\Md}{M_\mathrm{d}}

\newcommand{\phib}{\phi_\mathrm{bulge}}
\newcommand{\phid}{\phi_\mathrm{disc}}

\newcommand{\Nbody}{$N$-body\xspace}

\newcommand{\secref}[1]{Section~\ref{#1}}
\newcommand{\figref}[1]{Figure~\ref{#1}}
\newcommand{\tabref}[1]{Table~\ref{#1}}

\title[Metal-rich vs. metal-poor globular clusters in the Milky Way]{Origin of the metal-rich vs. metal-poor globular clusters dichotomies in the Milky Way: A sign of low black hole natal kicks}

\author[Rostami-Shirazi et al.]{
	Ali Rostami-Shirazi$^{1}$,
       Akram Hasani Zonoozi$^{1,2}$\thanks{E-mail: 
        a.hasani@iasbs.ac.ir},   
	Hosein Haghi$^{1,2,3}$, 
	and Malihe Rabiee$^{1}$ 
	\\
	$^{1}$Department of Physics, Institute for Advanced Studies in Basic Sciences (IASBS), PO Box 11365-9161, Zanjan, Iran\\
 $^{2}$Helmholtz-Institut f\"ur Strahlen-und Kernphysik (HISKP), Universit\"at Bonn, Nussallee 14-16, D-53115 Bonn, Germany\\
$^{3}$School of Astronomy, Institute for Research in Fundamental Sciences (IPM), PO Box 19395-5531, Tehran, Iran\\
}

\date{Accepted XXX. Received YYY; in original form ZZZ}

\pubyear{2024}

\begin{document}
\label{firstpage}
\pagerange{\pageref{firstpage}--\pageref{lastpage}}
\maketitle

\begin{abstract}
The bimodal metallicity distribution of globular clusters (GCs) in massive galaxies implies two distinct sub-populations: metal-poor and metal-rich. Using the recent data of \textit{Gaia} we highlighted three distinct dissimilarities between metal-poor and metal-rich GCs in the Milky Way (MW). Half-mass (light) radii of metal-poor GCs exhibit, on average, $\simeq 52 \pm$5 ($60 \pm$3) per cent more expansion than metal-rich ones. Furthermore, the lack of metal-poor GCs at low Galactocentric distances ($R_\mathrm{G}$) follows a triangular pattern in $R_\mathrm{G}$-[Fe/H] space, indicating that GCs with lower metallicities appear further away from the Galactic center. Metal-poor GCs are more susceptible to destruction by the tidal field in the inner part of the MW. We perform a series of \Nbody simulations of star clusters, to study the impact of the BHs' natal kicks on the long-term evolution of low- and high-metallicity GCs to explain these observational aspects. We found that the retention of BHs inside the cluster is crucial to reproducing the observed dissimilarities. The heavier and less expanded BH sub-system (BHSub) in metal-poor clusters leads to more intense few-body encounters, injecting more kinetic energy into the stellar population. Consequently, they experience larger expansion and higher evaporation rates rather than metal-rich clusters. The higher energy production within the BHSub of metal-poor GCs causes them to dissolve before a Hubble time near the Galactic center, leading to a triangular pattern in $R_\mathrm{G}$-[Fe/H] space.
\end{abstract}


\section{Introduction}\label{sec:intro}

The Globular clusters (GCs) found in most early-type galaxies, such as the Milky Way (MW), exhibit a bimodal distribution in terms of their metallicity and optical color \citep{Zinn1985,Forbes1997,Larsen2001,Bica2006,Peng2006,Brodie2006,Chies-Santos2012} indicating the presence of two distinct GC sub-populations: one with a bluer color and lower metal content, and the other with a redder color and higher metal content. The metal-rich and metal-poor GCs have systematically different locations and kinematics in their host galaxies.

Metal-rich GCs in galaxies are typically concentrated towards the center and have a radial distribution profile similar to the galaxy's spheroidal stellar component. Conversely, metal-poor GCs have a wider distribution and are more likely to be associated with the stellar halo \citep{Bassino2006,Goudfrooij2007,Peng2008,Strader2011,Forbes2012,Pota2013}. In the MW, the distribution of metal-rich GCs drops off sharply beyond a Galactocentric distance of about $10\kpc$. However, the population of metal-poor GCs extends much farther, covering a range from approximately 1 to $100\kpc$. Furthermore, the kinematics of the metal-rich sub-population are similar to those of the main stellar component, including rotation, while the metal-poor sub-population exhibits a larger velocity dispersion and little or no net rotation \citep{Brodie2006,Schuberth2010,Strader2011,Pota2013}. In terms of age, metal-rich GCs are estimated to be around two billion years younger on average than metal-poor ones \citep{Hansen2013}.

Moreover, the blue GCs in the MW and other nearby galaxies tend to have a larger half-light radius (by 17–30 per cent) compared to red GCs \citep{Kundu1998,jordan2004,jordan2005,Harris2009,Woodley2010,Strader2012}. Notably, this size disparity persists regardless of galactocentric distance \citep{Harris2009}. Earlier studies attributed this phenomenon to projection effects and the influence of the tidal field \citep{Larsenn2001}, as well as the combined effect of mass segregation and the dependence of main-sequence (MS) lifetimes on metallicity \citep{jordan2004,jordan2005}. However, \citet{Webb2012} later challenged the interpretation regarding projection effects. They suggested that the different spatial distributions of red and blue sub-populations do not significantly contribute to the observed size difference, either for the entire population or at specific galactocentric distances. Instead, \citet{Webb2012} proposed that the size discrepancy likely originates from differences in formation processes or subsequent evolution. In addition, the size difference between red and blue GCs is likely due to the different orbital distributions \citep{Webb2016}.

Several scenarios were proposed to explain the color bimodality of GC populations. According to the gas-rich merger scenario, early mergers between smaller galaxies at high redshifts result in the formation of metal-poor GCs, and later mergers involving more evolved galaxies in high-density environments lead to the formation of metal-rich GCs \citep{kravtsov2005,muratov2010}. In another scenario, it is suggested that the bimodal distribution of GCs does not necessarily result from two distinct epochs or modes of GC formation, but can instead be attributed to a galaxy's assembly history without invoking mergers as the primary mechanism for GC formation. In this scenario, a sub-population of metal-rich GCs forms alongside the bulk of the galaxy's stellar component during an early, intense dissipative phase. During a slower phase, the sub-population metal-poor GCs is thought to have been accreted through minor mergers or stripping from satellite galaxies \citep{Forbes1997_1,Forbes1997,cote1998,cote2000,Forbes2011,Arnold2011,Tonini2013}. Unlike the gas-rich merger scenario, this model proposes that GCs of different metallicities are formed in different galaxies and then brought together, rather than being formed within the same galaxy at different stages of its evolution. The findings related to the possibility of GCs escaping from dwarf galaxies and being captured by host galaxies confirm this scenario \citep{LAW2010b, Myeong1018,kipelman2019,massari2019, belazini2020, Kruijssen2020, Malhan2022, Rostami2022, Shirazi2023}.

The differences in characteristics between metal-poor and metal-rich GCs, in addition to shedding light on the formation history of GCs within galaxies, can provide valuable insights into the internal evolutionary pathways of GCs. Indeed, some of the observed disparities can be attributed to the clusters' internal evolution. For example, the presence of black holes (BHs) can significantly impact the dynamical evolution of GCs and cause more expansion in metal-poor clusters compared to metal-rich ones \citep{Downing2012,Mapelli2013,Banerjee2017,Debatri2022,DSC}. The extent of natal velocity kicks that BHs receive during their formation in a supernova explosion is still a matter of debate, leading to uncertainties about the initial number of BHs remaining in star clusters. It was initially believed that BHs experience a significant natal kick during their supernova explosion, which could accelerate them over the escape velocity, resulting in almost all BHs leaving the cluster as soon as they are formed. However, over the last decade, several observations have provided evidence of the possible presence of BHs in some GCs, suggesting that the natal kick received by BHs is not typically as strong as previously thought.
\citep{Maccarone2007,Shih2010,Barnard2011,Strader2012BH,Chomiuk2013,Giesers2018,Giesers2019,Saracino2022,Saracino2023}.  

On the theoretical and computational side, several studies show that the retention of BHs within certain clusters is necessary to reproduce their observable parameters \citep{banerjee2011,Peuten2016,Baumgardt2019,Zocchi2019,Gieles2021,Torniamenti2023}. Recent studies indicate that a BH mass fraction of 0-1 per cent of the MW GCs' present-day total mass is necessary to explain the observations \citep{Askar2018,Weatherford2018,Weatherford2020,Dickson2023}. In \citet{DSC}, utilizing direct $N$-body simulations, we highlighted that high BH natal kicks are not necessary to achieve this fraction; rather, even with high initial retention of BHs, a substantial number of them are depleted through few-body encounters, shaping the present-day BH mass fraction of MW GCs (see also \citealt{Ghasemi2024}).

In this paper, we aim to investigate how the natal kick received by BHs affects the observational parameters that distinguish between metal-poor and metal-rich GCs. Specifically, utilizing $N$-body calculations we evaluate whether the dissimilarities in the evolution of simulated metal-poor and metal-rich clusters are consistent with the observational evidence. The two scenarios we consider are: 1) BHs experience low natal kick which leads to the presence of a segregated BH sub-system (BHSub), or 2) BHs are ejected from the cluster immediately after their formation through a significant natal kick.  The paper is organized as follows: \secref{sec:observations}, describes the observational database used in this study. In \secref{sec:methodology}, we describe the initial setup of the $N$-body models. The main results are presented in \secref{sec:results} with a focus on the observational features of metal-poor and metal-rich simulated clusters in the presence (\secref{sec:Kick}) and absence (\secref{sec:NKick}) of BHs' natal velocity kick. Finally, \secref{sec:conclusion} provides a summary and conclusion.

\section{Observational Data}\label{sec:observations}

We compare the results of \Nbody simulations with the observed data of 151 Galactic GCs derived by fitting \Nbody models to an up-to-date compilation of ground-based radial velocities, $Gaia$ DR3 proper motions, and HST-based stellar mass functions \footnote{https://people.smp.uq.edu.au/HolgerBaumgardt/globular/}. We demonstrate three distinct differences between metal-poor and metal-rich GCs in the MW based on the observational properties including the orbital parameters \citep{Baumgardt2019mean}, projected half-light radius ($r_\mathrm{hl}$), 3D half-mass radius ($r_\mathrm{hm}$), the density inside the $r_\mathrm{hm}$ ($\rho$) \citep{Baumgardt2018}, and age \citep{Kruijssen2019}. We categorized GCs into metal-poor (blue) and metal-rich (red) sub-populations based on their $\mathrm{[Fe/H]}$ values \citep{Kruijssen2019,Harris2010}, using a threshold of $\mathrm{[Fe/H]}=-1.4$.

\begin{figure*}
	\centering
	\includegraphics[scale = 0.68]{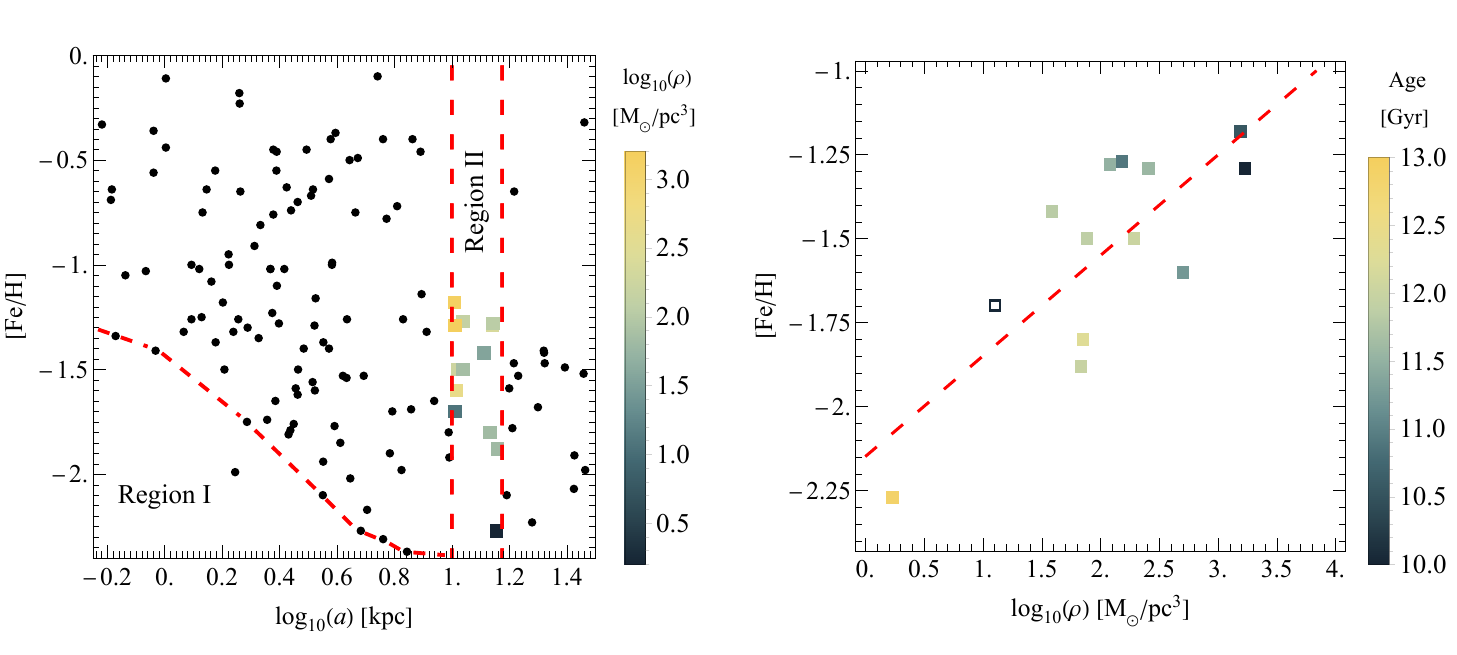}  	
	\caption{Left: the distribution of the MW GCs population in metallicity-semi-major axes space. Right: metallicity and density inside the $r_\mathrm{hm}$ of the GCs located in Region II. The colour codes correspond to density inside the $r_\mathrm{hm}$ (left panel) and age (right panel) of the GCs in Region II. The red dashed lines in the left panel indicate the boundaries of Regions I and II.}  
	\label{fig:observation-data}
\end{figure*}

\begin{figure}

  \centering\includegraphics[scale=0.36]{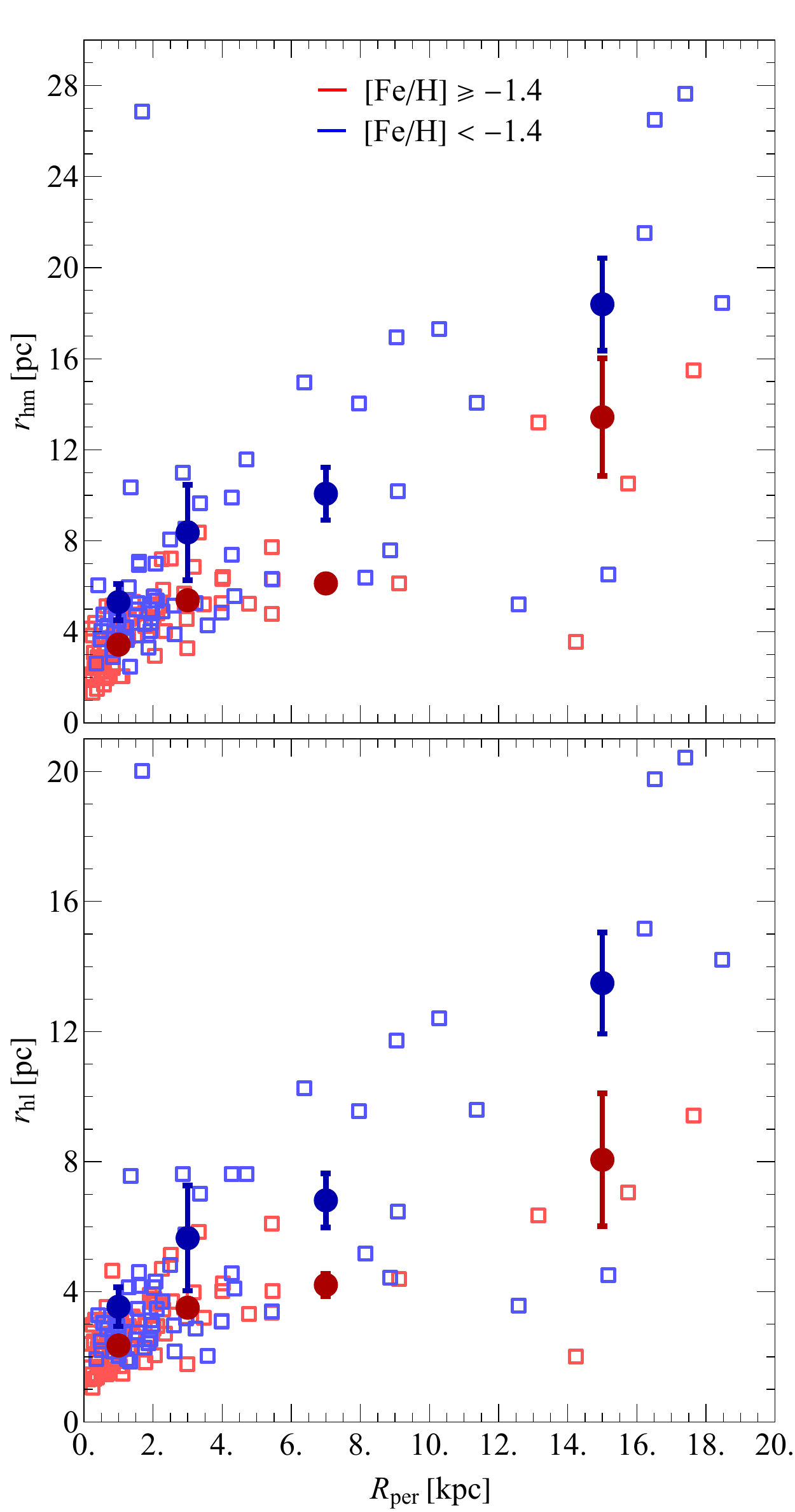}
  \caption{$r_\mathrm{hm}$ (top panel) and $r_\mathrm{hl}$ (bottom panel) of metal-poor (blue squares) and metal-rich (red squares) MW GCs vs. their perigalactic distances, $R_\mathrm{per}$. The blue and red circles respectively denote the mean size of the metal-poor and metal-rich GCs across four distinct $R_\mathrm{per}$ intervals.}
  \label{fig:rg}

\end{figure}

\subsection{The distribution of the MW GCs in $\mathrm{[Fe/H]}-a$ plane}\label{sec:region}

The metallicity distribution function of the MW GCs has peaked at around $\mathrm{[Fe/H]}\sim -0.5$  and $\mathrm{[Fe/H]}\sim -1.6$ for metal-rich and metal-poor GCs, respectively \citep{Harris2016}. Observational data indicates that GC populations of different metallicities have different radial distributions in the MW. The metallicity of the MW GCs is plotted versus their mean Galactocentric distances (semi-major axis, $a$) in the left panel of \figref{fig:observation-data}. As can be seen, a triangular region (Region I) appears in the [Fe/H]-$a$ diagram, corresponding to the inner part of the MW ($a<10\kpc$), and is devoid of metal-poor GCs. Region I exhibits a distinct pattern of a lack of metal-poor GCs at low Galactocentric distances ($R_\mathrm{G}$): GCs with lower metallicity appear farther away from the center of the Galaxy.

To compare the internal properties of GCs regardless of the impact of the tidal field, we evaluate the GCs that are distributed at approximately the same mean Galactocentric distances (e.g., $10\kpc \leq a < 15\kpc$), which include both metal-poor and metal-rich GCs (shown as Region II in \figref{fig:observation-data}). As can be seen, the density of GCs increases with metallicity such that the most metal-rich cluster is almost two orders of magnitude denser than the metal-poor GCs. This indicates that metal-poor GCs are less gravitationally bound compared to metal-rich ones, making them more vulnerable to tidal stripping and dissolution. This may provide insight into the absence of such GCs in Region I. The vulnerability of metal-poor GCs to tidal forces at the Galactic center may explain the lack of such GCs in Region I. The same analyses can be done for larger mean Galactocentric distances (e.g. $15\kpc\leq a$). However, in the inner part of the Galaxy, additional external effects such as rapid variation of tidal force with $R_\mathrm{G}$ \citep{Gnedin1997}, disk crossing tidal shocks \citep{Gnedin1999}, giant molecular clouds \citep{Gieles2006}, and spiral arms \citep{Gieles2007} may play an essential role in the present-day density of clusters. Regions I and II exhibit a fundamental difference in the evolution of metal-poor and metal-rich GCs. Tidal stripping is more likely to disrupt GCs with lower metallicity, especially in regions with strong tidal fields. That's why the survival of GCs with very low metallicity is only possible at the semi-major axis beyond $\sim 8\kpc$. We can propose this scenario: metal-poor GCs were initially distributed throughout the Galaxy, including its inner regions, but the strong dependence of a cluster's evolution and lifespan on its metallicity gave rise to Regions I and II.

The colour coding in the right panel of \figref{fig:observation-data}, corresponds to the age of GCs located in Region II. The figure provides evidence that there is a negative correlation between the age of GCs and both their metallicity and density. On average, metal-poor GCs are approximately 2$\Gyrs$ older than their metal-rich counterparts. It is possible to hypothesize that the huge disparity in evolution between metal-poor and metal-rich GCs may not be necessary to explain Regions I and II. Instead, the difference in age between them could be responsible for their difference in density in Region II and the absence of metal-poor GCs in Region I. In \secref{sec:Com}, we will examine this hypothesis and show that the age discrepancy between metal-poor and metal-rich GCs is inadequate to justify the existence of Regions I and II.

\subsection{The $r_\mathrm{hm}$ and $r_\mathrm{hl}$ of red and blue GC sub-populations}\label{sec:mass-light}

\begin{table}
	\centering
 \caption{The average $r_\mathrm{hm}$ and $r_\mathrm{hl}$ of red and blue sub-populations in the MW GCs at different pericentric distances. The first column shows the range of $R_\mathrm{per}$ in $\kpc$. Columns 2 to 5 show the mean $r_\mathrm{hm}$ and $r_\mathrm{hl}$ for red and blue GCs. Blue GCs exhibit a mean $r_\mathrm{hm}$ that is $52\pm 5$ per cent larger than red GCs, while the mean $r_\mathrm{hl}$ of blue GCs exceeds that of red ones by $60\pm 3$ per cent.}
 
	\begin{tabular}{ccccc}
		\hline 
		\hline  
		$R_\mathrm{per}$ & $\langle r_\mathrm{hm}\rangle$(red) & $\langle r_\mathrm{hm}\rangle$(blue) & $\langle r_\mathrm{hl}\rangle$(red) & $\langle r_\mathrm{hl}\rangle$(blue) \\ 
	$[\kpc]$ & $[\pc]$ & $[\pc]$ & $[\pc]$ & $[\pc]$ 
		\\
		\hline 
            (0,2) & 3.44 $\pm$ 0.15 & 5.31 $\pm$ 0.79 & 2.35 $\pm$ 0.10 & 3.54 $\pm$ 0.60 \\
            (2,4) & 5.39 $\pm$ 0.34  & 8.37 $\pm$ 2.09  & 3.51 $\pm$ 0.24  & 5.65 $\pm$ 1.62  \\
            (4,10) & 6.13 $\pm$ 0.36  & 10.07 $\pm$ 1.16  & 4.21 $\pm$ 0.35  & 6.81 $\pm$ 0.83  \\
            (10,100) & 13.43 $\pm$ 2.58  & 18.39 $\pm$ 2.03  & 8.06 $\pm$ 2.04  & 13.49 $\pm$ 1.56  \\

		\hline
	\end{tabular}

	\label{tab:initial_conditions}
\end{table}

The present-day size of GCs is determined by three competing physical mechanisms: tidal stripping, stellar evolution, and two-body relaxation and few-body encounters within GCs. \figref{fig:rg}, shows the $r_\mathrm{hm}$ (top panel), and $r_\mathrm{hl}$ (bottom panel) of MW GCs versus their perigalactic distance ($R_\mathrm{per}$). Considering all GCs in the MW, the average sizes of blue GCs are 94 per cent larger in terms of their half-mass radius ($r_\mathrm{hm}$) and 105 Per cent larger in terms of their half-light radius ($r_\mathrm{hl}$) compared to red GCs, regardless of their Galactocentric distances. Since blue GCs have a broader spatial distribution in the MW, a significant portion of this size disparity can be attributed to the impact of tidal stripping. To examine the internal dynamical evolution's impact on cluster radii without the effects of tidal stripping, it's essential to compare their sizes within the same intervals of Galactic radii. Therefore, to determine the pure effect of stellar evolution and internal dynamical evolution, we calculated the mean $r_\mathrm{hm}$ and $r_\mathrm{hl}$ of metal-poor and metal-rich GCs at four different intervals of $R_\mathrm{per}$. The bin sizes are adjusted to ensure the same number of GCs in each bin. The blue and red circles represent the mean size of metal-poor and metal-rich GCs, respectively, for the same $R_\mathrm{per}$ intervals.

The range of $R_\mathrm{per}$ and the corresponding mean $r_\mathrm{hm}$ and $r_\mathrm{hl}$ radius of metal-poor and metal-rich GCs are listed in  \tabref{tab:initial_conditions}. Taking into account all of the $R_\mathrm{per}$ intervals, we find that, the $r_\mathrm{hm}$($r_\mathrm{hl}$) of the metal-poor GCs sub-population is, on average, $52\pm 5$($60\pm 3$) per cent larger than that of the metal-rich ones. Since clusters, especially tidally filling ones, experience larger mass loss at $R_\mathrm{per}$, we expect $R_\mathrm{per}$ to have a more important effect on the dynamical evolution of the cluster, its radius, and dissolution time. However, choosing the intervals based on the $R_\mathrm{G}$, similar to \citet{Harris2009}, the difference between the $r_\mathrm{hm}$ and $r_\mathrm{hl}$ of the metal-poor and metal-rich GCs decrease to $25$ and $30$ per cent respectively.

The assembly scenario posits that metal-poor GCs, which form one of the Gaussian distributions, are accreted from satellite galaxies. This prompts an examination of whether the changing galactic potential during accretion could be responsible for the observed size discrepancy between metal-poor and metal-rich GCs. However, $N$-body simulations in time-dependent galactic potentials by \citet{Miholics2016} have shown that accreted clusters quickly adapt their sizes to match those of clusters born in the MW on the same orbit. Their findings suggest that, assuming similar initial sizes, clusters born in the Galaxy and those that are accreted cannot be separated based solely on their current size. Consequently, the observed size difference of approximately 50 per cent between metal-poor and metal-rich sub-populations can be attributed primarily to differences in their internal evolutionary pathways, rather than their origin or accretion history.

\begin{figure*}
	\centering
	\includegraphics[trim= 0 0 0 30,clip,scale = 0.59]{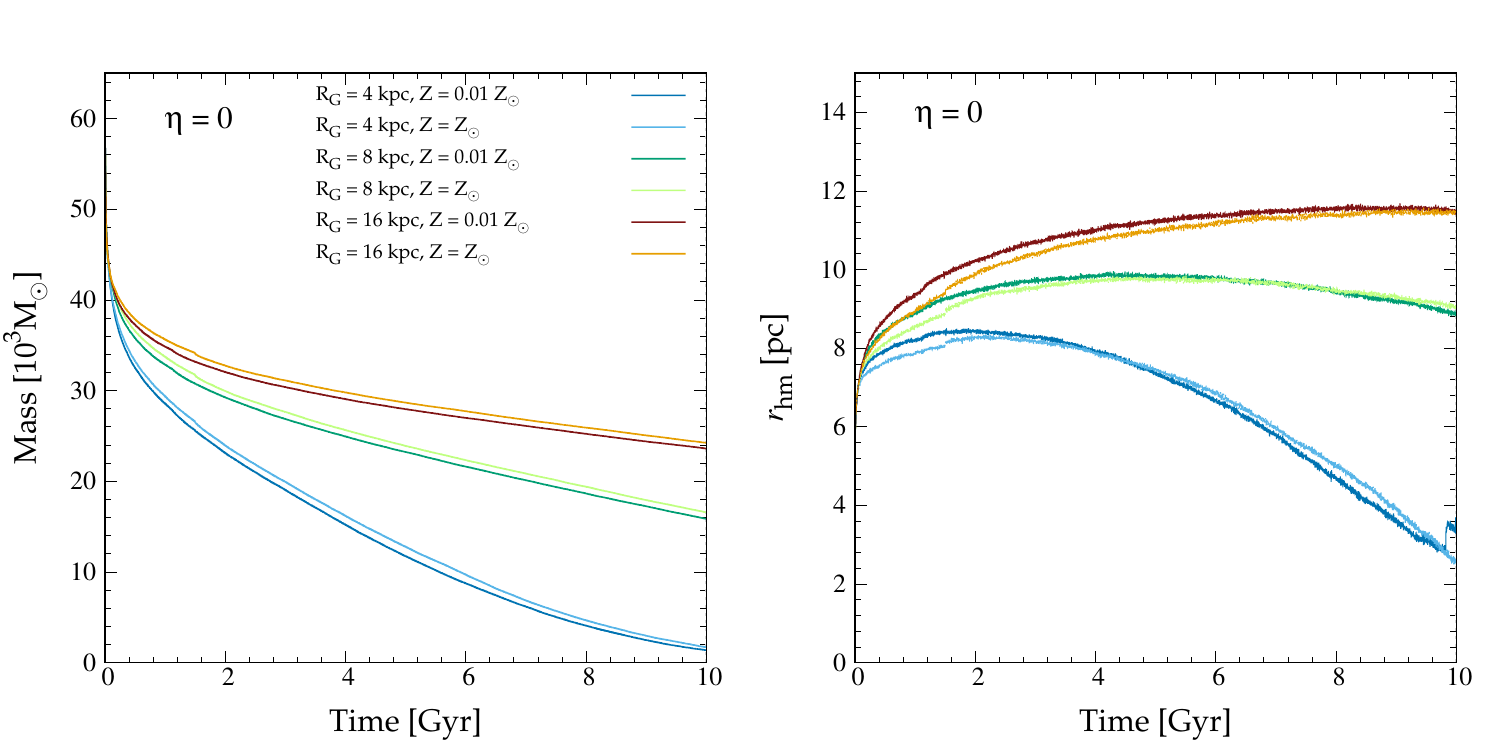}  	
	\caption{The evolution of the total mass (left panel) and $r_\mathrm{hm}$ (right panel) of the clusters with different $Z$ and $R_\mathrm{G}$. The dark lines indicate clusters with $Z=0.01Z_{\odot}$, while the light lines represent clusters with $Z=Z_{\odot}$. In all models, BHs are ejected from the cluster upon formation due to kick velocity ($\eta=0$).} 
	\label{fig:mass_rh_r0}
\end{figure*}

\section{DESCRIPTION OF THE MODELS AND INITIAL CONDITIONS}\label{sec:methodology}

To investigate the effect of metallicity ($Z$) on the evolution of star clusters with different retention fractions of BHs, we performed direct $N$-body simulations using the \textsc{NBODY7} code \citep{Aarseth2012}. The clusters are populated with stars in the mass range of $m_{\text{low}}=0.08\Msun$ and $m_{\text{up}} = 150\Msun$  following the canonical IMF \citep{Kroupa2001,Kroupa2013} and are initially in virial equilibrium and spatially distributed according to the Plummer density profile \citep{Plummer1911}.

We do not consider primordial binary stars for the sake of simplicity, although binaries can form via three-body interactions throughout the simulations. All models evolve up to 10 $\Gyr$. The initial mass and $r_\mathrm{hm}$ of all models are $M=6\times10^4 \Msun$ and $r_\mathrm{hm}= 5\pc$, respectively. We assume metallicities in the $Z = 0.01-1 Z_\odot$ range to cover metal-rich and metal-poor GCs. We place our modeled clusters on circular orbits at $R_\mathrm{G}=4-16\kpc$. The retention fraction, $\eta$, is also adopted as a free parameter to study how the amount of BHs remaining in a cluster can affect the long-term evolution of clusters with different metallicities. We consider two conditions for each model: 1) After cluster formation, all  BHs receive a kick velocity drawn from a Maxwellian distribution with a one-dimensional dispersion of $\sigma_{\text{kick}} = 190 \kms$ \citep{Hansen1997}, meaning that almost all BHs are kicked out of the cluster immediately after formation ($\eta=0$). 2) BHs do not receive any kick velocity and all of them are initially retained in the cluster ($\eta =1$).

We used a three-component Milky Way-like tidal field which is made up of a central bulge, a disc, and a phantom(\footnote{Phantom represents either the unconfirmed dark matter in the $\Lambda$CDM cosmological model, or the purely mathematical source of the Milgromian potential in Modified Newtonian Dynamics (MOND; \cite{Milgrom1983}).}) dark matter halo potential that is scaled so the circular velocity at $8.5\kpc$ is $220\kms$. The bulge is modeled as a central point mass:
\begin{equation}\label{eq:phi_bulge}
	\phib  = {\frac{-G\Mb}{R}},  	
\end{equation}
where $ M_\mathrm{b}=1.5\times10^{10}\Msun$ is the mass of the bulge component and $R=\sqrt{x^{2} + y^{2} + z^{2}}$ is the Galactocentric distance.

\begin{figure*}
	\includegraphics[trim= 0 0 0 0,clip,scale = 0.41]{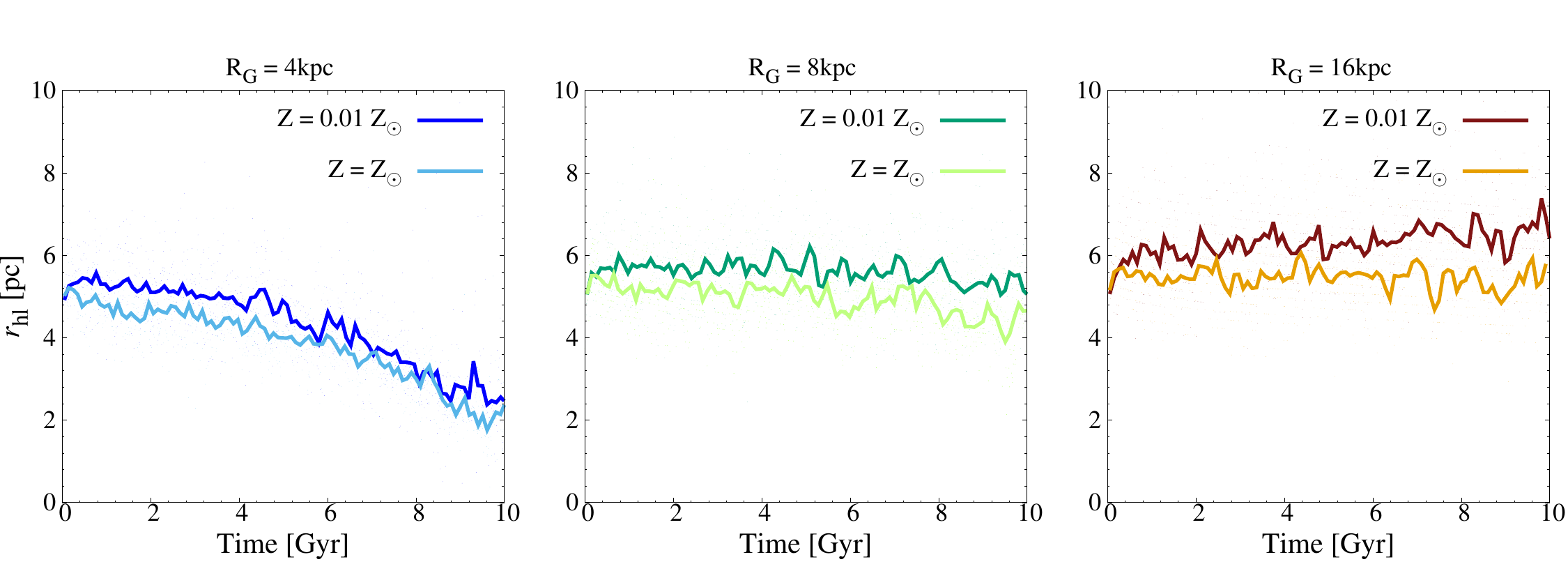}  	
	\caption{The evolution of the $r_\mathrm{hl}$ of models with $\eta=0$ and different metallicities evolving at different orbits with Galactocentric distance of $R_\mathrm{G}=4\kpc$ (blue), $8\kpc$ (green) and $16\kpc$ (brown). Dots show the $r_\mathrm{hl}$ in each snapshot, and the solid lines denote the mean $r_\mathrm{hl}$ per 100 Myrs.} 
	\label{fig:mass_rh_r1_1Gyr} 
\end{figure*}

The potential of  the disc is approximated by the  Miyamoto \& Nagai potential \citep{Miyamoto1975};
\begin{equation}\label{eq:miyamoto}
	\phid = {\frac{-G\Md}{\sqrt{x^2 + y^2 + \left(a+\sqrt{b^2 + z^2}\right)^2}} }
\end{equation}
where $a$ and $b$ are the disc scale length and height respectively, and $ M_\mathrm{d}$ is the total mass of the disc component. We used values of $a =4\kpc$ and $b=0.5\kpc $ for scale length and height \citep{Read2006}. The mass of the disc is adopted  to be  $M_\mathrm{d}=5.0\times10^{10}\Msun$ as suggested by \citet{Xue2008}. The dark matter halo is represented by a logarithmic potential given by:
\begin{equation}
        \phi_\mathrm{halo}= \frac{1}{2} V_\mathrm{circ}^2 \ \mathrm{ln} (R^2+R_\mathrm{circ}^2).
	\label{eq:3-3}
\end{equation}
The constant parameter $R_\mathrm{circ}$ is chosen such that the collective potential of the bulge, disc, and halo yields a circular velocity of $V_\mathrm{circ}=220\kms$ in the disc plane at a distance of $8.5\kpc$ from the Galactic centre.

\section{Results}\label{sec:results}

We conducted two sets of models with (\secref{sec:Kick}) and without (\secref{sec:NKick}) BH natal kick velocities to examine how the retention of BHs affects the evolution of metal-poor and metal-rich clusters. In section \ref{sec:Com}, we will assess whether the differences in the evolution of simulated metal-poor ($Z = 0.01 Z_\odot$) and metal-rich ($Z = Z_\odot$) clusters in each set ($\eta=0$ and $\eta=1$) are consistent with observational evidence.

\subsection{The presence of natal velocity kicks in BHs}\label{sec:Kick}

The evolutionary path of massive stars in the color-magnitude diagram can be significantly altered by metallicity, up until the formation of the final remnant. Massive metal-rich stars lose more mass through stellar winds than metal-poor stars \citep{Vink2001,vink2005,Belczynski2010}. As a result, BHs left over from metal-poor progenitor stars are expected to be heavier. BHs with a mass of up to $\sim 80 \Msun$ can form in metal-poor clusters, whereas the maximum mass of BHs in a solar metallicity cluster is $\sim 20 \Msun$ (\citealt{Belczynski2008, Belczynski2010}; also see Figure 1 in \citealt{Banerjee2017}). Moreover, the range of masses that convert to BHs is larger in metal-poor clusters \citep{Shanahan2015}. Therefore both the mass fraction and the average mass of BHs formed in a metal-poor cluster are always expected to be higher than that in a metal-rich cluster \citep{Hurley2000}. Generally, stars with lower metallicity evolve faster in the MS phase than those with higher metallicity \citep{Pols1998,Hurley2000}. This indicates that clusters of the same age have a higher MS turnoff mass in metal-rich clusters. According to the fitting formulae by \citet{Hurley2000}, the luminosity of a metal-rich MS star with a mass $<15\Msun$ is fainter than that of a metal-poor star. Therefore, the low-metallicity star has a lower mass-to-light ratio than the metal-rich star.

\begin{figure*}
	\centering
	\includegraphics[trim= 0 0 0 30,clip,scale = 0.62]{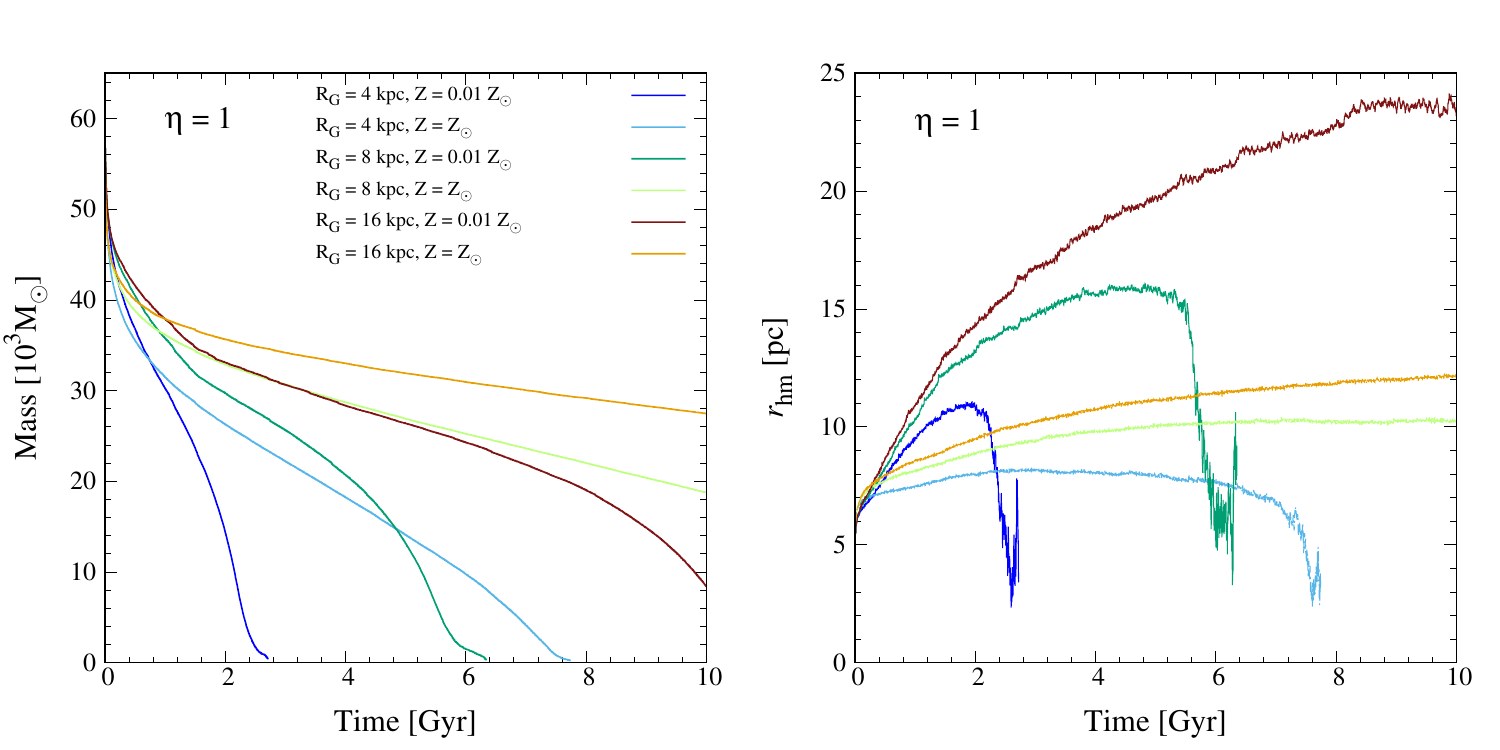}  	
	\caption{Same as \figref{fig:mass_rh_r0}, but for the simulated models with the presence of BH remnants ($\eta=1$).}  
	\label{fig:mass_rh_r1}
\end{figure*}

\begin{figure*}
	\includegraphics[trim= 0 0 0 0,clip,scale = 0.43]{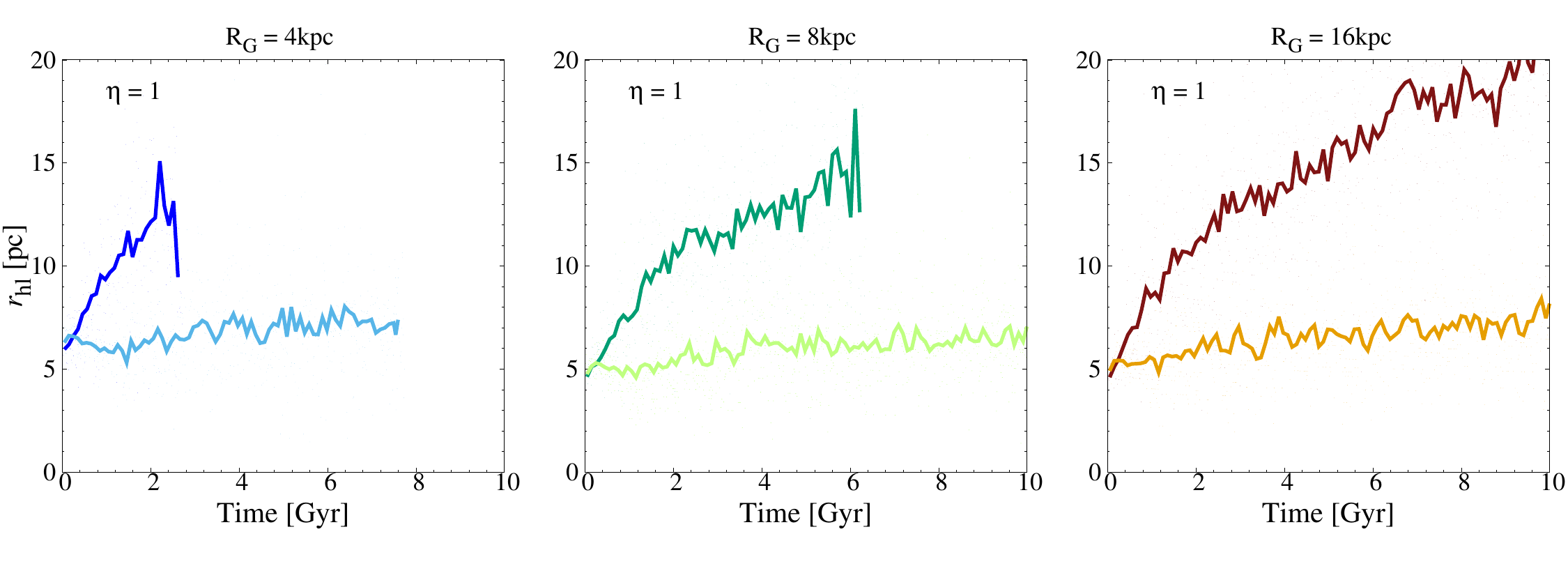}  	
	\caption{Same as Figure \ref{fig:mass_rh_r1_1Gyr}, but for models with $\eta=1$.}  
	\label{fig:mass_rh_r1_2Gyr}
\end{figure*}

\begin{figure}
	\includegraphics[trim= 0 0 0 30,clip,scale = 0.43]{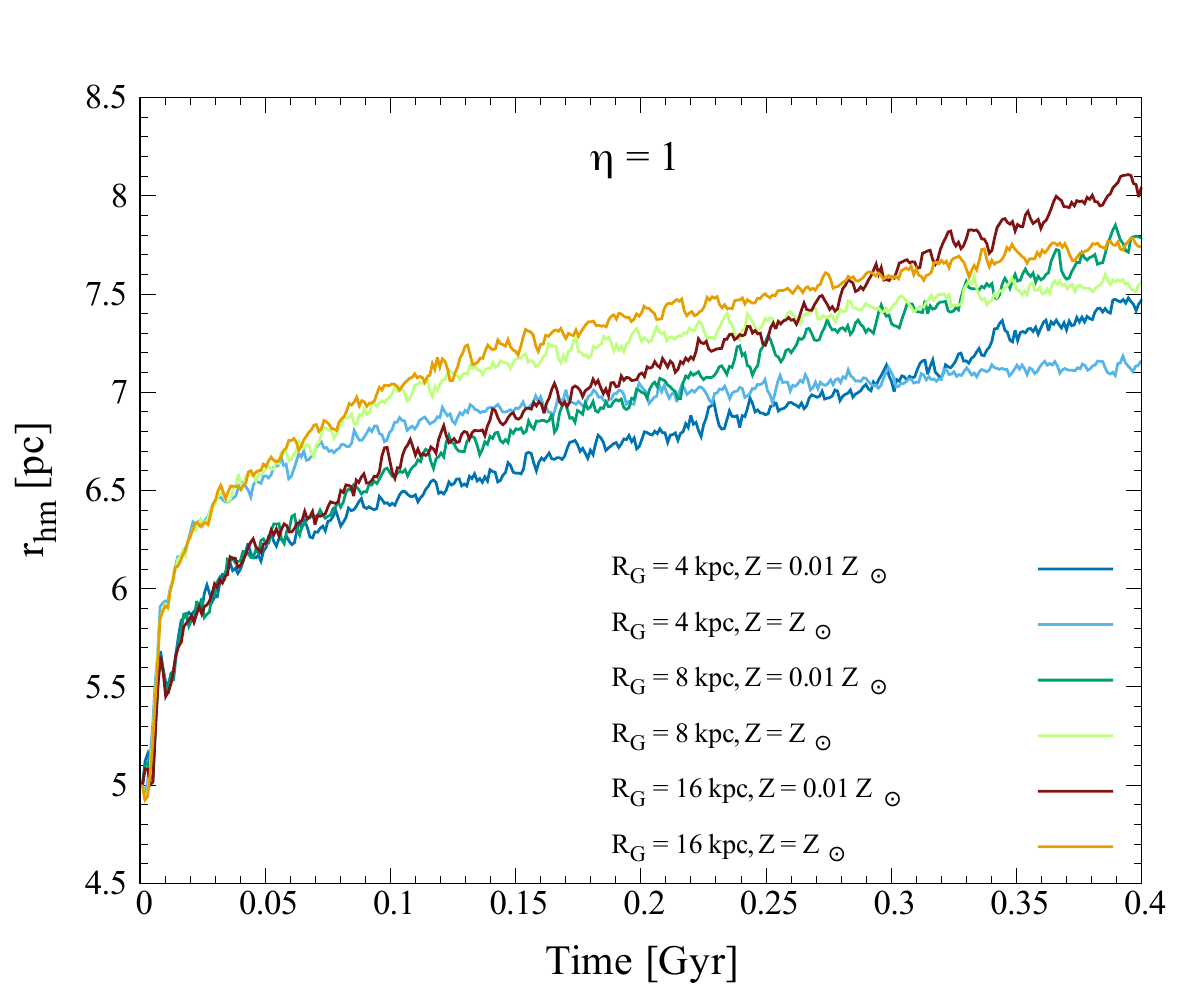}  	
	\caption{The evolution of the $r_\mathrm{hm}$ over the first 400 Myrs, for modeled clusters with  $\eta =1$ at different orbital radius ($R_\mathrm{G}=$ 4, 8, and 16 $\kpc$), with two distinct metallicities represented by dark ($Z=0.01Z_\odot$) and light ($Z=Z_\odot$) coloured lines.}  
	\label{fig:mass_rh_r1_first}
\end{figure}

\begin{figure*}
	\centering
	\includegraphics[trim= 0 0 0 0,clip,scale = 0.55]{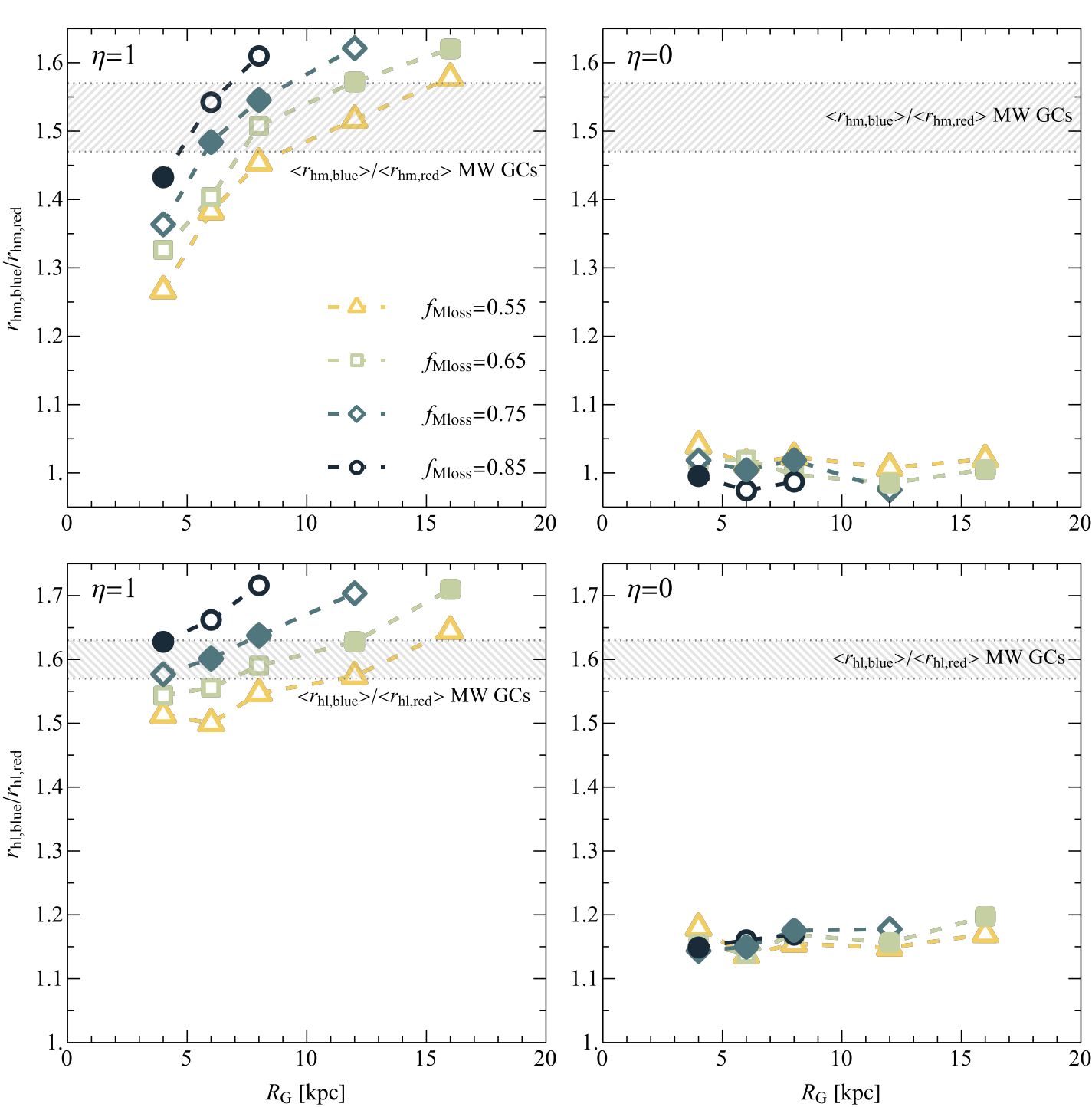}
  \caption{The ratio of $r_\mathrm{h}$ ($r_\mathrm{hl}$ and $r_\mathrm{hm}$) of blue ($Z=0.01Z_{\odot}$) to red ($Z=Z_{\odot}$) modeled clusters ($r_\mathrm{h,blue}/r_\mathrm{h,red}$) vs. Galactocentric distances for all simulated models at snapshots corresponding to mass loss fractions, $f_\mathrm{Mloss}$, of 0.55, 0.65, 0.75, and 0.85, without ($\eta =0$, right panels) and with ($\eta=1$, left panels) BH retention. Filled markers denote $r_\mathrm{h,blue}/r_\mathrm{h,red}$ ratios of simulated models at corresponding $f_\mathrm{Mloss}$ values of MW GCs at the same Galactic radii. According to \tabref{tab:initial_conditions}, the $r_\mathrm{hm}$ ($r_\mathrm{hl}$) of blue MW GCs is on average $52\pm 5$ ($60\pm 3$) per cent larger than that of red GCs, as indicated by the dashed areas.}
    \label{fig:GCs-Sim-rh}
\end{figure*}

\begin{figure*}
	\centering
	\includegraphics[scale = 0.55]{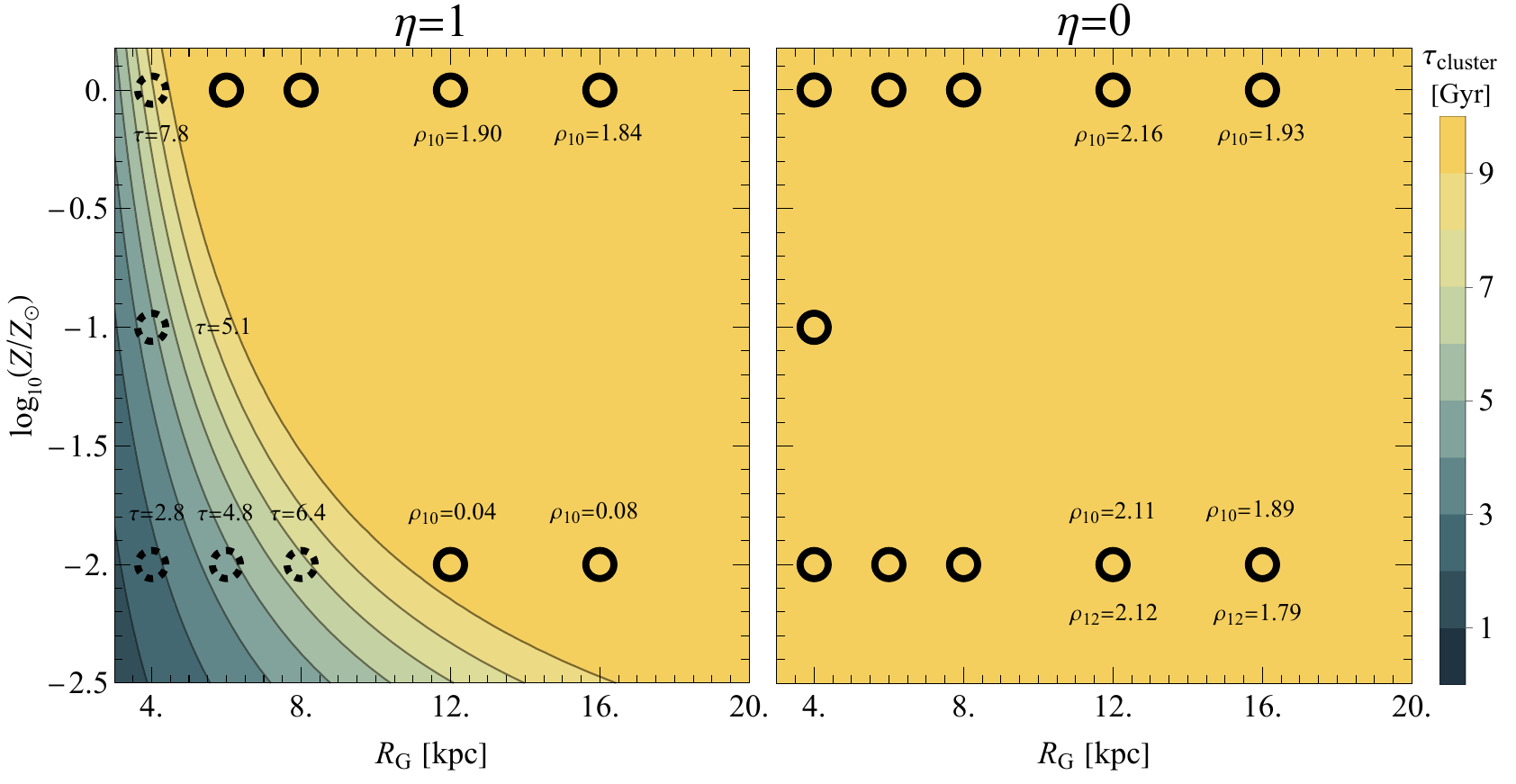}  	
        \caption{Galactocentric distances and metallicity of all 22 simulated clusters. The left and right panels correspond to simulated models with $\eta =1$ and $\eta =0$, respectively. The colour codes represent the dissolution time of the clusters ($\tau_\mathrm{cluster}$) obtained from the optimal fit. We set the maximum of the colour scale at $\tau_\mathrm{cluster}=10\Gyr$. The yellow areas indicate regions where clusters have survived for more than $10\Gyr$. The black circles represent simulated clusters with $\tau_\mathrm{cluster}> 10\Gyr$. The dashed line circles correspond to models that have dissolved before 10$\Gyr$. The panels display the dissolution times of all 22 models, as well as the density inside the $r_\mathrm{hm}$ of models with $R_\mathrm{G} > 10\kpc$ at $t=10 ~ (12) \Gyr$ ($\rho_\mathrm{10 ~ (12)} \ [\mathrm{\Msun/{pc}^3}]$).} 
	\label{fig:lifetime}
\end{figure*}

Assuming a high natal velocity kick for BHs ($\sigma_{\text{kick}} = 190 \kms$), we set up a series of models in which almost all BHs are kicked out from the cluster within the first $100 \ \mathrm{Myr}$ ($\eta =0$). \figref{fig:mass_rh_r0} shows the time evolution of the total mass and $r_\mathrm{hm}$ for modeled clusters with $\eta=0$.  At all $R_\mathrm{G}$, sub-solar metallicity clusters display only a slightly higher mass-loss compared to the solar metallicity clusters (left panel of \figref{fig:mass_rh_r0}). Moreover, metal-poor GCs reach a slightly larger maximum size, $r_\mathrm{hm}(\mathrm{max})$, compared to their metal-rich counterparts (right panel of \figref{fig:mass_rh_r0}). This is because a higher mass fraction of BHs forms in metal-poor clusters, and blue clusters will lose slightly more mass by ejecting BHs in the presence of natal kicks. Therefore, when clusters are emptied of BHs by high natal kicks, the metallicity has no significant effect on the dissolution time, $r_\mathrm{hm}$, and other mass-weighted radius.

The overall evolution of the $r_\mathrm{hl}$ is shown in \figref{fig:mass_rh_r1_1Gyr} for blue and red GCs located at 4, 8, and 16$\kpc$. Unlike the $r_\mathrm{hm}$, which evolves similarly in blue and red clusters, the $r_\mathrm{hl}$ are larger in blue clusters. Our results show that with $\eta=0$, the metal-poor clusters can achieve a $\sim15$ per cent larger $r_\mathrm{hl}$ on average compared to metal-rich clusters (\figref{fig:mass_rh_r1_1Gyr}). This result agrees with the findings of \citet{Sippel2012}, who suggested that the difference in $r_\mathrm{hl}$ between blue and red clusters is likely due to the interplay between stellar evolution and dynamical effects like mass segregation. In metal-rich clusters, more luminous mass is contained in MS stars due to the higher MS turnoff mass. Since mass segregation causes the most massive stars to sink to the central part of the clusters, the presence of less massive remnants in metal-rich clusters creates additional space in the core for MS stars. Hence, luminous stars steepen the luminosity profile in the central regions of the metal-rich clusters. On the other hand, the higher luminosity of MS stars in metal-poor clusters contributes to their overall brighter appearance beyond the central regions. \citet{Sippel2012} found that these two factors are responsible for size differences of up to approximately 20 per cent between clusters with high and low metallicity.
\subsection{The absence of natal velocity kicks in BHs }\label{sec:NKick}

When a large number of BHs remain in the cluster ($\eta =1$), they cannot reach energy equilibrium with low-mass stars through energy equipartition and consequently undergo runaway segregation toward the cluster center leading to the formation of a BHSub in the central part of the cluster \citep{Spitzer1987}. The centrally segregated BHSub exhibits significant dynamical activity, leading to the formation of numerous BH-BH binaries (BBHs) through three-body interactions \citep{Spitzer1987, Heggie2003}. Through the subsequent encounters between BBHs and single BHs, the binary systems experience gravitational tightening as individual BH companions scatter onto higher orbits. Consequently, this scattering mechanism imparts gained kinetic energy to the entire stellar population via two-body interactions, which leads to the expansion of the cluster \citep{Mackey2007,Mackey2008,Banerjee2017}. The scattered BH returns to the cluster center via dynamical friction. Each such encounter hardens the BBH and increases the recoil velocity. As a result, BHSub operates akin to an energetic power plant that pumps kinetic energy from the cluster's central region to its surrounding stellar population.

The core radius of metal-rich clusters experiences a greater expansion compared to that of metal-poor clusters, as mass-loss by stellar winds in metal-rich clusters removes more matter from the core. As a result, metal-poor clusters have a higher density of BHSubs and a higher frequency of three-body collisions\citep{Mapelli2013}. On the other hand, the energy produced from dynamical interactions in BHSubs is intensified by increasing the number of BHs and their mean mass \citep{DSC}. Consequently, in metal-poor clusters, energy injection from BHSub is larger, leading to greater expansion and higher evaporation rates.

\figref{fig:mass_rh_r1} shows the time evolution of total mass and $r_\mathrm{hm}$ for different simulated models with $\eta=1$ at $R_\mathrm{G}=4, 8$ and $16\kpc$. The mass fraction of BHs formed in a metal-poor cluster with $Z=0.01 Z_\odot$ is about $f_\mathrm{BH}=6.6$ per cent, which is three times larger than that of a metal-rich cluster ($Z=Z_\odot$) with $f_\mathrm{BH}=2.2$ per cent. This difference in the mass fractions of BHs leads to a clear gap in the evolution of their $r_\mathrm{hm}$ and also in their dissolution time. The comparison between the right panel of Figures \ref{fig:mass_rh_r0} and \ref{fig:mass_rh_r1} shows that the observed enhancement in $r_\mathrm{hm}$ of blue clusters is mostly due to the dynamical activity of BHs. Furthermore, due to the presence of gravitationally hotter BHSub, blue clusters experience a higher evaporation rate and, as a result, dissolve earlier compared to red clusters (left panel of \figref{fig:mass_rh_r1}).

In \figref{fig:mass_rh_r1_first}, the evolution of the $r_\mathrm{hm}$ over the first 400 Myr is shown for modeled clusters with $\eta =1$.  Clusters with solar metallicity are observed to expand faster in the first $\sim300$ Myr compared to clusters with sub-solar metallicity. This difference can be attributed to the fact that during the early stages of cluster evolution, stellar evolution plays a dominant role compared to dynamical heating caused by three-body encounters within the BHSub \citep{Mapelli2013,trani2014}. At this stage, solar metallicity clusters lose more mass through stellar winds and supernova expulsion, leading to more expansion compared to sub-solar metallicity clusters. However, this trend reverses after $\sim300$ Myr, and dynamical heating becomes the dominant process, resulting in more expansion in sub-solar metallicity clusters.

The $r_\mathrm{hl}$ are clearly larger in blue clusters for both sets of $\eta=0$ and $\eta=1$ models, regardless of whether $r_\mathrm{hm}$ is larger or not (Figures \ref{fig:mass_rh_r1_1Gyr} and \ref{fig:mass_rh_r1_2Gyr}). The more expansion in the $r_\mathrm{hm}$ of blue clusters will enhance any differences in $r_\mathrm{hl}$ produced by differences in the luminosity function. \figref{fig:mass_rh_r1_2Gyr} shows a substantial difference in $r_\mathrm{hl}$ between blue and red clusters in models with $\eta=1$ compared to models with $\eta=0$.

\subsection{Comparing with observations}\label{sec:Com}

In this section, we determine whether the presence or absence of BH natal kicks can accurately reproduce the observed discrepancy between the metal-poor and metal-rich sub-populations of MW GCs (see Sec. 2).
\subsubsection{The evolution of half-mass (light) radii}

In \figref{fig:GCs-Sim-rh} the ratio of radii of blue to red clusters for two categories of $\eta=$1 (left panels) and 0 (right panels) are compared with observation. As the MW GCs have lost approximately 60-90 per cent of their initial stellar masses through stellar evolution or long-term dynamical evolution \citep{Webb2015,Baumgardt2017,Baumgardt2019mean}, $r_\mathrm{hl}$ and $r_\mathrm{hm}$ have been determined at snapshots corresponding to mass loss fractions, $f_\mathrm{Mloss}$, of 0.55, 0.65, 0.75, and 0.85. Tidal stripping, more pronounced near the Galactic centre, leads to varying mass loss fractions in GCs. MW GCs, on average, experienced $f_\mathrm{Mloss}$ of $0.85$, $0.72$, and $0.66$ for GCs orbiting at $a\leq5\kpc$, $5<a\leq10\kpc$, and $a>10\kpc$, respectively\footnote{https://people.smp.uq.edu.au/HolgerBaumgardt/globular/}. To ensure our simulations accurately reflect the observed MW GC population, we analyze the size ratio of blue to red clusters at mass loss fractions corresponding to their Galactic radii. Specifically, we use $f_\mathrm{Mloss} = 0.85$, $0.75$, and $0.65$ for simulated models at $\RG = 4\kpc$, $\RG \in \{6, 8\}\kpc$, and $\RG \in \{12, 16\}\kpc$, respectively. These values are denoted in \figref{fig:GCs-Sim-rh} by filled markers.

If BHs receive natal kicks that cause them to eject from the cluster ($\eta=0$), then there is no significant difference in the $r_\mathrm{hm}$ of blue and red clusters for all $\RG$ and $f_\mathrm{Mloss}$, which is consistent with previous studies \citep{Downing2012,Sippel2012}. However, the $r_\mathrm{hl}$ of blue clusters is, on average, 15 per cent larger than that of red clusters which is in good agreement with the result of \citet{Sippel2012}. As can be seen in the right panels of \figref{fig:GCs-Sim-rh}, the modeled clusters with no BH retention ($\eta=0$) are unable to reproduce the observed values of the ratio of radii of blue to red sub-populations of MW GCs.

In contrast, keeping BHs in the cluster results in approximately 50-70 per cent greater expansion in metal-poor clusters than metal-rich ones (left panels of \figref{fig:GCs-Sim-rh}). The tidal radius limits the expansion of tidally filling clusters. Consequently, the stronger expansion of metal-poor clusters, driven by additional energy generated from their BHSub, is less constrained by the host galaxy's tidal field in the outer regions. As a result, the ratio of $\frac{r_\mathrm{h,blue}}{r_\mathrm{h,red}}$ increases with $R_\mathrm{G}$. As clusters experience progressive mass loss, the energy production from their BHSub increasingly dominates their evolution. Consequently, at higher $f_\mathrm{Mloss}$, the additional energy produced in metal-poor clusters amplifies the disparity in $r_\mathrm{hm}$ and $r_\mathrm{hl}$ between blue and red sub-populations, resulting in elevated $\frac{r_\mathrm{h,blue}}{r_\mathrm{h,red}}$ ratios (left panels of \figref{fig:GCs-Sim-rh}). The greater expansion in $r_\mathrm{hm}$ for blue clusters will exacerbate any discrepancies in $r_\mathrm{hl}$ produced by differences in the luminosity function. As illustrated in the bottom left panel of \figref{fig:GCs-Sim-rh}, the difference in $r_\mathrm{hl}$ between blue and red clusters exceeds their $r_\mathrm{hm}$ difference.

The left panels of \figref{fig:GCs-Sim-rh} demonstrate that the $\frac{r_\mathrm{h,blue}}{r_\mathrm{h,red}}$ ratios of simulated models for various $R_\mathrm{G}$, at corresponding $f_\mathrm{Mloss}$ values of MW GCs at the same Galactic radii (indicated by filled markers), align closely with the observed ratios in MW GCs (represented by the dashed area). As a result, adopting a low BHs natal velocity kick can nearly reproduce the observed $\sim50$ and $\sim60$ per cent disparity in the ratios of $\frac{r_\mathrm{hm,blue}}{r_\mathrm{hm,red}}$ and  $\frac{r_\mathrm{hl,blue}}{r_\mathrm{hl,red}}$, respectively. We should note that by choosing intervals based on the $R_\mathrm{G}$ instead of $R_\mathrm{per}$, the average $<r_\mathrm{hm}>$ ($<r_\mathrm{hl}>$) of blue clusters to red ones reduces to about 1.25 (1.3), which is still unreachable for $\eta=0$ models but can probably be achieved for models with  BHs retention fraction less than 100 per cent.


\subsubsection{Dissolution time}

Metallicities and Galactocentric distances of all 22 modeled clusters are shown in  \figref{fig:lifetime}. By determining the lifetime of each model, we calculated the optimal fit for the dissolution time ($\tau_\mathrm{cluster}$) of these clusters as a function of $R_\mathrm{G}$ and $Z$. The colour coding and corresponding curves indicate the lifetime of clusters with different $R_\mathrm{G}$ and $Z$. The uniform yellow color in the right panel of \figref{fig:lifetime} implies that the dissolution time of models with $\eta=0$ exceeds $10\Gyr$. This indicates that both metal-rich and metal-poor clusters survive in the Galaxy. In other words, the disruption rate of star clusters does not depend on their metallicity if no BHs or a few BHs are retained in the cluster. The survival of metal-poor clusters within the inner part of the MW leads to the absence of Region I in the $R_\mathrm{G}$-$Z$ space. This is also evident from the density of metal-rich and metal-poor clusters at $10\Gyr$, $\rho_{10}$, which is remarkably similar at the same $R_\mathrm{G}$. This is in contrast to the observed two-order of-magnitude disparity in density between metal-poor and metal-rich clusters in Region II (\figref{fig:observation-data}).

Since metal-poor GCs are, on average, $2 \Gyr$ older than their metal-rich counterparts \citep{Hansen2013}, one might hypothesize that extending the evolutionary time of the blue clusters up to $12 \Gyr$ would lead to their dissolution within the inner part of the MW, thereby resulting in the emergence of the Region I. Furthermore, there might be a distinct discrepancy in the density between blue and red clusters within Region II. Extending the simulations of modeled clusters with $Z=0.01 Z_\odot$ up to $12 \Gyr$ demonstrates that these clusters can survive at all $R_\mathrm{G}$ ($\tau_\mathrm{cluster}>12\Gyr$). Furthermore, in Region II, the density of these clusters remains approximately stable between the time intervals of $10$-$12 \Gyr$ ($\rho_{10} \approx \rho_{12}$). Therefore, despite the significant age gap of 2 billion years between the blue and red clusters, it remains impossible to reproduce the observational evidence for modeled clusters with $\eta=0$.

The BHSub in blue clusters is more massive and has a higher temperature, which results in more energy transfer to the outer stellar populations and more expansion. In the inner part of the Galaxy, where the escape velocity is lower, this extra energy can push stars beyond their escape velocities and significantly accelerate the evaporation rate of blue clusters. The left panel of \figref{fig:lifetime} shows the lifetime of clusters with $\eta=1$ in which the dissolution time strongly depends on their metallicity. The extra energy injection from BHSub in metal-poor clusters causes them to dissolve before 10 $\Gyr$ ($\tau_\mathrm{cluster}<10\Gyr$), in the inner part of the Galaxy ($R_\mathrm{G}<10\kpc$).

As the $R_\mathrm{G}$ increases, the clusters are affected by the weaker Galactic tidal field, which implies an increased escape velocity in these clusters. Therefore, the energy produced from the BHSub of metal-poor clusters has less influence on the evaporation rate and the dissolution time exceeds 10$\Gyr$ (left panel of \figref{fig:lifetime}). Indeed, the dissolution time of clusters in the inner part of the Galaxy strongly depends on their metallicity, if most of the BHs remain in clusters such that the metal-poor clusters located in the inner part of the Galaxy have shorter lifetimes ($\tau_\mathrm{cluster}<10\Gyr$) and mostly dissolve. These clusters are shown with dashed circles, while the surviving clusters are depicted with solid circles. Therefore, adopting a low natal velocity kick for BHs successfully explains the emergence of Region I in $Z$-$R_\mathrm{G}$ space. Furthermore, a substantial disparity of two orders of magnitude is observed when comparing the $\rho_{10}$ between solar and sub-solar metallicity modeled clusters at $R_\mathrm{G}=12$ and 16 $\kpc$ (Region II). Given that the impact of BHSub on the evaporation rate of clusters is influenced by the tidal field, the extent of Region I also depends on the tidal field of the host galaxy. In other words, for a host galaxy with a stronger tidal field than the MW, Region I would be wider.

For models with $\eta=1$ the difference in lifetime between metal-poor and metal-rich clusters is substantial, especially near the Galactic center. For instance, at a distance of 4 kpc, a cluster with metallicity $Z=Z_\odot$ survives almost three times longer than a cluster with metallicity $Z=0.01Z_\odot$ (the left panel of \figref{fig:lifetime}). By increasing the number of stars in the cluster, the size of the region where metal-poor clusters exist for less than 10 Gyr would likely decrease, but a triangular region (Region I) would remain. In our simulations with $10^5$ stars, clusters with a metallicity of $Z=0.01Z_\odot$ could not survive at $R_\mathrm{G}<12$ kpc for 10 Gyr, while clusters with $Z=Z_\odot$ at $R_\mathrm{G}=4.5$ kpc survived beyond 10 Gyr. However, to ensure that if the metal-poor clusters were an order of magnitude more massive, one would still see the gap in $R_\mathrm{G}-Z$ space, higher-$N$ simulations should be performed.

It should be noted that in reality, clusters have larger masses and smaller radii compared to our artificial models. Additionally, the initial BHs retention fraction has been constrained to be less than $100$ per cent by recent studies \citep{Peuten2016,Baumgardt2017}. These factors typically result in GCs having longer dissolution times compared to our results and shifting the $\tau_\mathrm{cluster}$ curves to smaller $R_\mathrm{G}$. BH natal kicks are, to date, poorly constrained and understood from both observational and theoretical point of view. A common model \citep{Belczynski2008, Fryer2012} for supernova natal kick magnitude assumes NS-like kicks \citep{Hobbs2005} for BHs as well, but which are scaled down linearly with an increasing material fallback fraction, so-called the canonical supernova kicks. In this formalism, For the most massive BHs that form via the direct collapse of a massive star, no natal kicks are imparted. Metal-poor clusters have more massive BHs (with masses larger than 40 $\mathrm{M_\odot}$) which are more likely to collapse directly to BH without supernova explosion. Therefore, high-mass BHs that are more populated in metal-poor clusters should receive no natal kick. Here we adopt the same natal kick for BHs in both metal-poor and metal-rich clusters. Hence in a more realistic simulation, this results in larger differences in the size and lifetime of metal-poor and metal-rich clusters.

\section{DISCUSSION AND Conclusions}\label{sec:conclusion}

The distribution of metallicity in GCs found in spiral and giant elliptical galaxies follows a bimodal pattern, indicating the presence of two distinct sub-populations - one metal-poor and the other metal-rich. These sub-populations differ in their kinematics, size, age, and spatial distribution within their host galaxies. We highlighted three distinct differences between metal-poor and metal-rich MW GCs: I-II) on average, the $r_\mathrm{hm}$ and $r_\mathrm{hl}$ of metal-poor GCs are approximately $\sim 52 \pm$5 and $\sim 60 \pm$3 per cent larger than those of metal-rich ones, respectively. III) There is an absence of metal-poor GCs in the inner part of the MW, which follows a triangular pattern in $R_\mathrm{G}$-[Fe/H] space (Region I; see \figref{fig:observation-data}). This pattern shows that GCs with lower metallicities appear further away from the Galactic center. Metal-poor GCs are more susceptible to destruction by the tidal field in proximity to the MW before a Hubble time. This provides evidence that the dissolution time of GCs in the inner part of the Galaxy strongly depends on their metallicity. To understand the cause of these differences, we performed direct $N$-body simulations to study the long-term evolution of the metal-poor and metal-rich GCs for two sets of star clusters with and without BHs natal velocity kicks. In both sets, we assessed whether the dissimilarity in the evolution of simulated metal-poor ($Z = 0.01 Z_\odot$) and metal-rich ($Z = Z_\odot$) clusters is consistent with the observational evidence. The main outcomes of our study can be summarized as follows:

\begin{itemize}
    \item In the presence of high natal velocity kick for BHs ($\eta=0$), the evolution of $r_\mathrm{hm}$ is nearly the same in both metal-poor and metal-rich clusters. However, due to differences in the luminosity function, the sub-solar metallicity clusters had a larger $r_\mathrm{hl}$ by approximately 15 per cent than solar metallicity clusters. These results are consistent with \citet{Sippel2012}, who found that when BHs are ejected from clusters by high natal kicks, the $r_\mathrm{hl}$ of metal-poor clusters appear, on average, 17 per cent larger than that of metal-rich clusters. However, they found no significant difference between the two sub-populations in the $r_\mathrm{hm}$. Moreover, our simulations reveal that both metal-rich and metal-poor clusters exhibit nearly identical dissolution times. This implies that no empty region of metal-poor clusters in the inner part of the MW is expected. Therefore, the scenario of high natal velocity kicks for BHs fails to explain the observed dichotomies between metal-poor and metal-rich GCs.

   \item The absence of high natal velocity kicks for BHs ($\eta=1$) leads to the formation of BHSubs in the central part of the cluster. Since the BHSubs of metal-poor clusters are heavier and less expanded than those of metal-rich clusters, few-body encounters are more intense in the BHSubs of metal-poor clusters, resulting in greater injection of kinetic energy into the entire stellar population. Consequently, metal-poor clusters experience larger expansion and higher evaporation rates compared to their metal-rich counterparts (in agreement with \citealt{Downing2012,Mapelli2013,Banerjee2017,Debatri2022,DSC}). The $r_\mathrm{hl}$ and $r_\mathrm{hm}$ of metal-poor clusters evolve to approximately 50-70 per cent larger than metal-rich ones. Moreover, sub-solar metallicity clusters situated in proximity to the Galaxy dissolved before a Hubble time. This leads to the emergence of empty regions of metal-poor clusters in $R_\mathrm{G}$-$Z$ space. Hence, adopting a low natal kick velocity for BHs can successfully overcome all three observational evidence.

   \item  Given the observed triangular void region of metal-poor GCs in the $R_\mathrm{G}$-[Fe/H] parameter space, we propose a scenario to elucidate the present-day radial distribution of metal-poor GCs: metal-poor GCs were initially distributed throughout the Galaxy, including its inner regions, but the strong dependence of the dynamical evolution of star clusters on metallicity leads to the rapid dissolution of metal-poor GCs in the inner part of the MW and their absence in Region I. Our simulations demonstrate that assuming a low natal velocity kick for BHs, there is a strong correlation between the dissolution time of clusters and their metallicity, particularly in the inner part of the host galaxy. To test this scenario, it's important to conduct observational investigations with two main objectives. First, to determine whether the metal-poor-empty triangular region observed in the $R_\mathrm{g}$-[Fe/H] parameter space is a universal pattern in early-type galaxies. Second, to examine how the morphology of this void region is influenced by the mass$/$Luminosity of the host galaxies (Rostam Shirazi et al. in prep.). 

   \item By simulating GCs with different retention fractions, it is possible to narrow down the fraction of BHs that remain in the cluster by comparing the observed metal-poor-empty triangular region in $R_\mathrm{G}$-[Fe/H] space with the dissolution time of modeled clusters.
    
\end{itemize}
\section*{Data availability}
The data underlying this article are available in the article.


\bibliographystyle{mnras}
\bibliography{references}

\appendix

\bsp
\label{lastpage}
\end{document}